\documentclass[sigconf,review]{acmart}

\AtBeginDocument{%
  }

\copyrightyear{2024}
\acmYear{2024}
\setcopyright{acmlicensed}\acmConference[ASE '24]{39th IEEE/ACM International Conference on Automated Software Engineering }{October 27-November 1, 2024}{Sacramento, CA, USA}
\acmBooktitle{39th IEEE/ACM International Conference on Automated Software Engineering (ASE '24), October 27-November 1, 2024, Sacramento, CA, USA}
\acmDOI{10.1145/3691620.3695344}
\acmISBN{979-8-4007-1248-7/24/10}

\usepackage{amsmath,amsfonts}
\usepackage[linesnumbered,ruled,vlined]{algorithm2e}
\usepackage{pifont}
\usepackage{amsthm}
\usepackage{algpseudocode}
\usepackage{tcolorbox}
\usepackage{array}
\usepackage[caption=false,font=normalsize,labelfont=sf,textfont=sf]{subfig}
\usepackage{textcomp}
\usepackage{stfloats}
\usepackage{url}
\usepackage{verbatim}
\usepackage{graphicx}
\usepackage{ulem}
\usepackage{amsfonts}
\usepackage[utf8]{inputenc}
\usepackage{setspace}
\usepackage{cleveref}
\crefname{section}{§}{§§}
\Crefname{section}{§}{§§}
\usepackage{caption}
\usepackage{booktabs}
\usepackage{makecell}
\usepackage{MnSymbol}
\usepackage{utfsym}
\usepackage{filecontents, pgffor}
\usepackage{balance}
\usepackage{latexsym}
\usepackage{color,xcolor,colortbl}
\usepackage{multirow}

\definecolor{mySourceNodecolor}{HTML}{ddf1e8}
\definecolor{myOtherNodecolor}{HTML}{385775}
 \usepackage{tikz}
\usepackage{xurl}

\begin{document}

\title{Detecting Malicious Accounts in Web3 through Transaction Graph}

\author{Wenkai Li}
\affiliation{%
  \institution{Hainan University}
  \city{Haikou}
  \country{China}}
\email{liwenkai871@gmail.com}

\author{Zhijie Liu}
\affiliation{%
  \institution{ShanghaiTech University}
  \city{Shanghai}
  \country{China}}
\email{liuzhj2022@shanghaitech.edu.cn}

\author{Xiaoqi Li}
\affiliation{%
  \institution{Hainan University}
  \city{Haikou}
  \country{China}}
\email{csxqli@gmail.com}
\authornote{Corresponding author}

\author{Sen Nie}
\affiliation{%
  \institution{Keen Security Lab, Tencent}
  \city{Shanghai}
  \country{China}}
\email{snie@tencent.com}


\begin{abstract}
The web3 applications have recently been growing, especially on the Ethereum platform, starting to become the target of scammers. The web3 scams, imitating the services provided by legitimate platforms, mimic regular activity to deceive users. The current phishing account detection tools utilize graph learning or sampling algorithms to obtain graph features. However, large-scale transaction networks with temporal attributes conform to a power-law distribution, posing challenges in detecting web3 scams. In this paper, we present ScamSweeper, a \textit{novel} framework to identify web3 scams on Ethereum. Furthermore, we collect a large-scale transaction dataset consisting of web3 scams, phishing, and normal accounts. Our experiments indicate that ScamSweeper exceeds the state-of-the-art in detecting web3 scams.
\end{abstract}


\maketitle
\pagestyle{empty}  
\thispagestyle{empty} 

\section{Introduction}
\label{sec:intro}

Detecting malicious accounts on the Ethereum network is a significant area of research~\cite{li2024defitail,li2024stateguard}. Due to its high transaction volume and anonymity, Ethereum has been a prime target for fraudulent activities~\cite{kong2024characterizing,niu2024unveiling,li2023overview,mao2024automated}. Numerous studies have focused on identifying phishing accounts to safeguard users from scams~\cite{grover2016node2vec,he2023txphishscope,li2022ttagn, perozzi2014deepwalk, poursafaei2021sigtran,kim2024drainclog}. However, with the emergence of Ethereum, web3 scams have also emerged, where certain services engage in covert malicious activities within the blockchain. These scams attract users and generate profits by concealing specific malicious activities within legitimate services, such as concealing scams in the NFT airdrops~\cite{NFTAirdrop2024}.

Web3 scams have been upgraded from traditional phishing scams on Ethereum~\cite{li2024stateguard,niu2024unveiling}. Under the chain, traditional phishing is similar to web3 scams, which trick users into connecting their wallets to fake mediums. On the chain, traditional phishing accounts defraud users' funds. However, the web3 scams pretend to offer regular services while allowing the attacker to gain access to the user's funds surreptitiously. Subsequently, attackers could pilfer tokens without the user's involvement and obscure their traces. Thus, 
one of the primary disparities lies in the mimic behavior and track obscuration by utilizing multiple transfers. 

In this paper, we present the ScamSweeper for identifying web3 scam accounts on Ethereum. Specifically, we enhance the sample walking methods to construct the structure and temporal-related transaction graphs. Then, the graph could be sliced into several directed subgraphs. After that, we sort the subgraphs as a sequence and feed them into a transposed Transformer to capture the dynamic evolution of subgraphs. Finally, we can detect the accounts.

\section{ScamSweeper}
\label{sec::method}

\subsection{Graph Construction} \label{sub:graph_construction}
Before learning the features of transaction graphs, we need to obtain the transactions and construct the transactions as a network graph. To alleviate the requirements of large-scale transaction data on computing resources, we implement Structure Temporal Random Walk (STRWalk) to sample the network. STRWalk can be divided into two steps. Firstly, given the current node $v_i$, it gets the neighbors $\{v_i^1,v_i^2,...,v_i^n\}$ of $v_i$, it computes the set of temporal attributes $T=\{t^1,t^2, ...,t^n\}$ for all the edges connected neighbors of $v_i$, where $t^j \in T$ means the timestamp value of edge between $v_i$ and $v_i^j$, $j \in [1,n]$. Then, the probability that each edge would be selected will be calculated. We randomly select a neighbor of $n_i$ according to the probability. Secondly, suppose that the current node $v_i$ has finished the first step of STRWalk, it selects the related nodes at the same temporal interval according to another probability. We repeat the first and second steps until the sampled walk sequence reaches a fixed length or the nodes have no neighbors. Finally, the output is a subgraph sequence $S = \{g_{v_0}^0,g_{v_0}^1,...,g_{v_0}^m\}$ ranked by the time, where $g_{v_0}^i$ is the subgraph at the $i_{th}$ temporal interval, $v_0$ represents the start node, and $m$ is the length of the sequence.

\subsection{Dynamic Evolution Learning}
\label{subsec:graph_encoder}



We consider that multiple edges can be between any pair of nodes, and each edge connects two nodes. To construct the edge feature matrix $\mathbb{E}$, we use a matrix $W \in \mathbb{R}^{1 \times n}$ to represent $\mathbb{E} = WE$. Moreover, the node feature matrix can be repressed by $\mathbb{V}= [WX_f, WX_t]^T$.

\noindent\textbf{Neighbor Alignment. }Due to the diverse neighbor numbers of each subgraph, the dimensions of neighborhood features $X^\mathcal{V}$ and directed edge features $X^\mathcal{E}$ are different, so we need to align the neighborhood features. We normalize the attribute features of the neighbor nodes with the $LeakyRelu$ activation function, and we also use the zero padding method to unify their dimensions.
\label{subsec:neighbor_alignment}

\noindent\textbf{Feature Extraction.}
\label{subsec:feature_extraction}
The subgraph is characterized by its node-centric structure, in which each constituent node in the neighborhood possesses a different significance to the central node. To quantify the neighborhood with different contributions, we employ the GAT~\cite{velivckovic2017graph} to delineate the behavior patterns. It facilitates calculating the importance degree among nodes and learning graph-pattern features, thereby learning the structure feature in each subgraph. By extracting the feature $h_g$ of each subgraph $G_g$ in all the temporal intervals sequentially, we can obtain the temporal subgraph feature sequence $\Phi=(h_g^0,h_g^1,...,h_g^m)\in \mathbb{R}^{m\times D}$ for \cref{subsec:sequence_learning}, where $D$ means the hidden size, $m$ is the variable temporal length.

\noindent\textbf{Feature Learning.}\label{subsec:sequence_learning}
To capture the dynamic evolution in the temporal subgraph feature sequence in \cref{subsec:graph_encoder}, we employ the transposed Transformer structure. First, we feed the time series into linear layers $\Theta_Q$, $\Theta_K$, and $\Theta_V \in \mathbb{R}^{d \times d}$, generating queries $Q$, keys $K$, and values $V\in \mathbb{R}^{m\times d}$. Then the self-attention and mask layers are employed.  Moreover, to enrich the feature, we implement the feed-forward layer, which is practically a two-layer fully connected network. Then, we employ a deep neural network~\cite{zeng2023financial} to discern temporal sequence relationships across different locations. Lastly, the Projection head~\cite{vaswani2017attention} is used to make the classification.

\section{Result}
We conduct a series of experiments to evaluate the ScamSweeper.

\noindent RQ1: \textbf{Is ScamSweeper effective for detecting Web3 scams?}\label{subsec:RQ3}

To validate the effectiveness of ScamSweeper in detecting web3 scams, we conduct experiments with various graph learning techniques (i.e., GAT~\cite{velivckovic2017graph}, GCN~\cite{kipf2016semi}, GraphSAGE~\cite{hamilton2017inductive}), and sequence learning method (i.e., Transformer~\cite{vaswani2017attention}). We collect a large-scale transaction dataset of web3 scam nodes for evaluation, two types of malicious nodes on the Ethereum blockchain. Leveraging the BlockchainSpider tool~\cite{wu2023tracer} with the depth-first algorithm, we scrape the first 18 million blocks. During this period, we get the dataset with the 3,125 transaction networks from the web3 scam Database~\cite{scamsniffer2024database}. After evaluation, we find that when the structural window sets 10, ScamSweeper achieves a weighted F1-score of 0.70, significantly higher than 17.29\%-48.94\% compared to other methods.

\noindent RQ2: \textbf{Is each component in ScamSweeper effective?} \label{subsec:RQ2}

To show the efficiency of the components (\cref{sub:graph_construction}, \cref{subsec:graph_encoder}, and \cref{subsec:sequence_learning}), we conduct an ablation experiment. In the subgraph learning stage (See \cref{subsec:graph_encoder}), we directly remove the graph-learning layer for comparison, illustrating the impact of graph-learning techniques within ScamSweeper. In the subgraph sequence learning phase (See \cref{subsec:sequence_learning}), we substitute the transposed Transformer encoder with a conventional Transformer structure, e.g., the different structure depicted in Figure \ref{fig:t-Transformer}, indicating the influence of the transposed Transformer. The experimental results depict that the graph and sequence learning components improve the F1-scores by 47.17\% and 90.24\%, respectively, with a structural window size of 5. 
Moreover, the weighted F1-scores are enhanced by 52.17\% and 27.27\%, respectively, with a structural window size of 10. 



\section{Conclusion}
\label{sec::conclusion}

This paper proposes ScamSweeper, a detection model for detecting mimic accounts in web3 scams on Ethereum. We utilize the structure temporal random walk to sample the node network, exploring the dynamic evolution of the subgraph sequence. After evaluating a real-world web3 scam dataset, ScamSweeper outperforms other methods by 15\% in weighted F1-score. Moreover, ScamSweeper has an advantage of 17.5\% in the F1-score on a phishing dataset.

\section{ACKNOWLEDGMENTS}
This work is sponsored by the National Natural Science Foundation of China (No.62362021 and No.62402146), CCF-Tencent Rhino-Bird Open Research Fund (No.RAGR20230115), and Hainan Provincial Department of Education Project (No.HNJG2023-10).

\normalem
\bibliographystyle{ACM-Reference-Format}
\bibliography{ref}

\balance

\end{document}